\documentclass[prb,preprint]{revtex4-1} 

\usepackage{amsmath}  
\usepackage{amsfonts} 
\usepackage{graphicx} 

\newcommand{\be}{\begin{equation}} 
\newcommand{\ee}{\end{equation}}

\begin{document}

\title{Three new roads to the Planck scale}
\author{Valerio Faraoni}
\email{vfaraoni@ubishops.ca} 
\affiliation{Department of Physics and Astronomy, Bishop's University,
2600 College Street, Sherbrooke, Qu\'ebec Canada J1M~1Z7}


\begin{abstract}

Three new heuristic derivations of the Planck scale are 
described. They are based on basic principles or phenomena 
of relativistic gravity and quantum physics. The Planck 
scale quantities thus obtained are within one order of 
magnitude of the ``standard'' ones. We contemplate the pair 
creation of causal bubbles so small that they can be 
treated as particles, the scattering of a matter wave off 
the background curvature of spacetime that it induces, and 
the Hawking evaporation of a black hole in a single burst 
at the Planck scale.

\end{abstract}

\maketitle 

\section{Introduction} 
\label{sec:1}

General relativity and quantum mechanics are two great 
achievements of twentieth century physics. Gravity is 
completely classical in Einstein's theory of general 
relativity, and quantum mechanics (broadly defined to 
include quantum field theory and particle physics) 
incorporates special relativity 
but excludes gravity. It is  believed that these two 
completely separate theories should merge at the Planck 
scale, at which general-relativistic effects become 
comparable to quantum ones. No definitive theory of quantum 
gravity is available, although much work has gone into 
string theories, loop quantum gravity, and other approaches 
({\em e.g.},\cite{1,2,3,4} see also \cite{5} and 
see \cite{6} for a popular exposition).

The Planck scale was introduced by Planck himself \cite{7} in 
1899, therefore predating the Planck law for blackbody 
radiation. The importance of the Planck units was realized 
by Eddington \cite{8} and the idea that 
gravitation and quantum 
mechanics should be taken into account simultaneously at 
this scale was spread by Wheeler \cite{9,10} and has 
bounced 
around ever since. The themes that a fundamental system 
of units exists in nature and that the values of these 
units can perhaps be derived in a super-theory have been 
the subject of a large literature (see 
Ref.~\cite{11} for an excellent introduction).

All derivations of the Planck scale more 
or less correspond to taking various combinations of the 
fundamental constants $G$ (Newton's constant) associated 
with gravity, $c$ (the speed of light {\em in vacuo}) 
characterizing relativity, and the Planck constant $h$ (or 
the reduced Planck constant $ \hbar \equiv h/(2\pi)$) which 
signals quantum mechanics. Usually the Planck scale is 
deduced, following Planck, on a purely dimensional 
basis \cite{7} or it is derived using the concept of 
a black 
hole in conjunction with that of a matter wave. The simplest 
derivation of the Planck scale notes that by combining the 
three fundamental constants $G,c$, and $\hbar$ one obtains 
a unique quantity with the dimensions of a length, the 
Planck length 
\be \label{Plength} 
l_\text{pl}= \sqrt{ 
\frac{G\hbar}{c^3}}=1.6 \cdot 10^{-35}\, \mbox{m} \,. 
\ee 
By combining $l_\text{pl}$ with $G$ and $c$ one then 
obtains the Planck mass 
\be \label{Pmass} 
m_\text{pl}=\frac{l_\text{pl}c^2}{G}= \sqrt{ \frac{\hbar 
c}{G}} = 2.2\cdot 10^{-8} \, \mbox{kg} \,, 
\ee 
the Planck energy 
\be E_\text{pl}= m_\text{pl} c^2= \sqrt{ \frac{\hbar 
c^5}{G}}= 1.3 \cdot 10^{19} \, \mbox{GeV} \,, 
\ee 
the Planck mass density 
\be \label{Pdensity} 
\rho_\text{pl} = 
\frac{m_\text{pl} }{l^3_\text{pl} } = 
\frac{c^2}{l_\text{pl}^2 G}= \frac{c^5}{\hbar G^2}=5.2\cdot 
10^{96} \, \mbox{kg} \cdot \mbox{m}^{-3} \,, 
\ee 
and the 
Planck temperature 
\be \label{Ptemperature} 
T_\text{pl}= 
\frac{E_\text{pl}}{k_\text{B}}= \frac{l_\text{pl} 
c^4}{k_\text{B} G} = 
\sqrt{ \frac{\hbar c^5}{G k_\text{B}^2}} =1.4 \cdot 10^{32} 
\, 
\mbox{K} \,, 
\ee 
where $k_\text{B}$ is the Boltzmann constant. We 
denote with $x_\text{pl}$ the Planck scale value of a 
quantity $x$ determined by dimensional analysis as in the 
above.  Two suggestive alternative derivations of the 
Planck scale appear in the literature and are reviewed in 
the next two subsections. At least six more roads to the 
Planck scale, which are slightly more complicated, are known and 
have been discussed in Ref.~[15]. How many ways to obtain the 
Planck scale without a full quantum gravity theory are possible? 
The challenge of finding them can be fun and very creative. Other 
possibilities to 
heuristically derive the Planck scale certainly exist: in 
Secs.~\ref{sec:2}-\ref{sec:4} we propose three new ones 
based on pair creation of ``particle-universes'', the 
propagation of matter waves on a curved  spacetime,   
or the Hawking radiation from black holes.

\subsection{A Planck size black hole}

In what is probably the most popular derivation of the 
Planck scale, one postulates that a particle of mass $m$ 
and Compton wavelength $\lambda=h/(mc)$, which has Planck 
energy, collapses to a black hole of radius $  
R_\text{S}=2Gm/c^2$ (the Schwarzschild radius of a 
spherical 
static black hole of mass $m$ \cite{12,13}). Like all orders of 
magnitude estimates, this procedure is not rigorous since it 
extrapolates the concepts of black hole and of Compton wavelength 
to a new regime in which both concepts would probably lose their 
accepted meanings and would, strictly speaking, cease being valid. 
However, this is how one gains intuition into a new physical 
regime. 

Equating the Compton wavelength of this mass $m$ to its black hole  
radius gives
\be 
m= \sqrt{ \frac{hc}{2G}} 
=\sqrt{\pi} \, m_\text{pl} \simeq 1.77\, m_\text{pl} \,. 
\ee

\subsection{A universe of size comparable with its Compton 
wavelength}

It is not compulsory to restrict to black holes in 
heuristic 
derivations of the Planck scale, although black holes 
certainly constitute some of the most characteristic 
phenomena predicted by relativistic gravity.\cite{12,13} Why not 
use a relativistic universe instead of a black hole? This 
approach is followed in the following argument  
proposed in John Barrow's  {\em Book of 
Universes} \cite{16} (but it does not appear in 
the technical literature and it definitely deserves to be included 
in the pedagogical literature).

Cosmology can only be described in a fully consistent and 
general way by a relativistic theory of gravity and one can 
rightly regard a description of the universe as 
phenomenology of relativistic gravity on par with the 
prediction of black holes. Consider a spatially homogeneous 
and isotropic universe which, for simplicity, will be taken 
to be a spatially flat 
Friedmann-Lema\^itre-Robertson-Walker spacetime with line 
element
\be 
ds^2=-dt^2 +a^2(t) \left( dx^2+dy^2+dz^2 \right) \,, 
\ee 
and with scale factor $a(t)$ and Hubble parameter $H(t) 
\equiv \dot{a}/a$. An overdot denotes differentiation 
with respect to the comoving time $t$ measured by observers 
who see the 3-space around them homogeneous and isotropic. 
The size of the observable universe is its Hubble radius $ 
cH^{-1}$ 
which is also, in order of magnitude, the 
radius of 
 curvature (in the sense of four-dimensional curvature) of 
this space. Consider the 
mass $m$ enclosed in a Hubble sphere, given by
\be 
mc^2 =\frac{4\pi}{3} \, \rho \left( H^{-1} c \right)^3= 
\frac{H^{-1} c^5}{2G} \,, \label{expressionofm} 
\ee 
where $\rho$ is the cosmological 
energy density and in the last equality we used the 
Friedmann equation \cite{12,13}  
\be\label{Friedmann} 
H^2=\frac{8 \pi G}{3c^2} \, 
\rho 
\ee 
(note that, following standard notation, $\rho_\text{pl}$ 
and $\rho$ denote a {\em mass density} and an {\em energy 
density}, respectively). 
The Planck scale is reached when the Compton wavelength 
of the mass $m$ is comparable with the Hubble radius, {\em 
i.e.}, when 
\be  \label{procedure}
\frac{c}{H} \sim \lambda=\frac{h}{mc} \,.
\ee 
This procedure implies that quantum effects (Compton 
wavelength) are of the same order of gravitational effects 
(cosmology described by the Friedmann equation). Clearly, we 
extrapolate Eq.~(\ref{Friedmann}) to a new quantum gravity 
regime 
from the realm of validity of general relativity and we 
extrapolate the concept of 
Compton wavelength from the realm of ordinary quantum mechanics. 
This extrapolation is necessary in order to learn something about 
the Planck scale, although it is not rigorous.

The expression~(\ref{expressionofm}) of $m$ then gives 
\be 
H^2=\frac{ c^5}{2Gh} \,. 
\ee 
Using again Eq.~(\ref{Friedmann}) yields the energy density
\be 
\rho \sim \frac{3c^7}{16\pi G^2 h} 
=\frac{3c^2}{32\pi^2} \, \rho_\text{pl} \simeq 10^{-2} 
c^2 \rho_\text{pl} \,, 
\ee
from which the other Planck 
quantities~(\ref{Plength})-(\ref{Ptemperature}) can be 
deduced by dimensional analysis. One obtains
\begin{eqnarray}
l &= & \frac{c}{\sqrt{G\rho}} \simeq 10 \, l_\text{pl} 
\,,\\ 
&&\nonumber\\
m &=& \frac{lc^2}{G} \simeq 10 \, m_\text{pl} \,,\\
&&\nonumber\\
E &=& mc^2 \simeq 10 \, E_\text{pl} \,,\\
&&\nonumber\\
T & = & \frac{E}{k_\text{B} } \simeq 10 \, T_\text{pl} \,.
\end{eqnarray}
At first sight the argument of a universe with size 
comparable with its 
Compton wavelength is not dissimilar in spirit from the 
popular 
argument comparing the Schwarzschild radius of a black hole 
with its Compton wavelength. In fact, it is commonly 
remarked that the universe is a relativistic system by 
showing that the size of the observable universe is the 
same as the Schwarzschild radius of the mass $m$ contained 
in it, for
\be
R_\text{S} =\frac{2Gm}{c^2} =\frac{2G}{c^2} \left( 
\frac{4\pi R^3}{3} \, \frac{\rho}{c^2} \right) = 
\frac{2G}{c^2} \, 
\frac{4\pi}{3} \, \frac{\rho}{c^2} \left( H^{-1}c \right)^3 
=\frac{8\pi G}{3c}\, H^{-3}\rho \,.
\ee
Equation~(\ref{Friedmann}) then yields $R_\text{S} \simeq 
cH^{-1}$, 
which is often reported in the popular science literature 
by 
saying that the universe is a giant black hole. 
This argument is definitely too naive because the 
Schwarzschild 
radius pertains to the Schwarzschild solution of the 
Einstein equations,\cite{12,13} which is very different from the 
Friedmann-Lema\^{i}tre-Robertson-Walker one. If 
one accepted this argument, then comparing the size of the 
visible universe $cH^{-1}$ with the Compton wavelength of 
the mass contained in it would be numerically similar  to 
comparing its Schwarzschild radius with 
this wavelength. However, the step describing the 
visible universe as a black hole (which is extremely  
questionable if not altogether incorrect) is logically not 
needed in the  procedure expressed by 
Eq.~(\ref{procedure}). 

Turning things around but in keeping with the spirit of the 
derivation above, it has also been noted that 
equating the Planck density to the density of a sphere 
containing the mass of the observable universe produces the size of 
the nucleus (or the pion Compton wavelength) as the radius of this 
sphere.\cite{20}

\section{Pair creation of particle-universes}
\label{sec:2}

Another approach to the Planck scale is the following. The idea of 
a universe which is quantum-mechanical in nature has been present 
in the literature for a long time and the use of the 
uncertainty principle to argue something about the universe 
goes back to Tryon's 1973 proposal that the universe may 
have originated as a vacuum fluctuation.\cite{Tryon} This notion of 
creation features prominently also in recent popular 
literature.\cite{Krauss} Consider now universes so small that they 
are 
 ruled by quantum mechanics and regard the 
mass-energies contained in them as  
elementary particles. At high energies there could be 
production of 
pairs of such ``particle-antiparticle universes''. Again, one goes 
beyond known and explored regimes of general relativity and 
ordinary quantum mechanics by extrapolating facts well known in 
these regimes to the unknown Planck regime. The 
Heisenberg uncertainty principle $\Delta E \Delta t \geq 
\hbar/2$ can be used by assuming that $\Delta E$ is the energy 
contained in a Friedmann-Lema\^itre-Robertson-Walker 
causal bubble of radius $R\sim H^{-1}c$ containing the 
energy $ 
\Delta E \simeq 4\pi\rho R^3  /3$. Setting $\Delta t 
\sim H^{-1}$ (the age of this very young universe), 
$\Delta E \Delta t \simeq \hbar/2$ gives
\be
\frac{4\pi}{3} \, \rho \left( H^{-1} c \right)^3 
H^{-1} \simeq \frac{\hbar}{2} 
\ee
which can be rewritten as 
\be
\frac{8\pi G}{3} \, \rho \, \frac{c^3}{G H^4}= \hbar \,.
\ee
Equation~(\ref{Friedmann}) then yields the mass density
\be
\frac{\rho}{c^2} \simeq \frac{3c^5}{8\pi G^2 \hbar} = 
\frac{3}{8\pi} \, \rho_\text{pl} \,, 
\ee 
one order of magnitude smaller than the ``standard'' Planck 
mass density~(\ref{Pdensity}). The 
other Planckian quantities can then be derived from $\rho$ 
and the fundamental constants $G, c$, and $h$.

\section{Scattering of a matter wave off the background 
curvature of spacetime} 
\label{sec:3}

The second alternative road to the Planck scale comes from 
the fact that, in general, waves propagating on a curved 
background spacetime scatter off it.\cite{DWB, 17, 18, 19} This 
phenomenon is 
well known and can be interpreted as if these waves had an 
effective mass induced by the spacetime curvature. It is 
experienced by waves with wavelength $\lambda$ 
comparable with, or larger than, the radius of curvature 
$L$ of 
spacetime. High frequency waves do not ``feel'' the 
larger scale inhomogeneities of the spacetime curvature 
and, as is intuitive, essentially propagate as if they 
were in flat 
spacetime.\cite{17, 18, 12, 19}  The phenomenon is 
not dissimilar 
from the scattering experienced by a wave propagating 
through an inhomogeneous medium when its wavelength is 
comparable with the typical size of the inhomogeneities. 
Again, we extrapolate the backscattering 
of a test-field wave by the (fixed) background curvature of 
spacetime to a new regime in which this wave packet gravitates, 
bends spacetime and, at the Planck scale,  impedes its own 
propagation.  Clearly, this extrapolation is not rigorous, like all 
order of magnitude estimates. However, we can gain some confidence 
in this procedure {\em a posteriori} by noting that it produces a 
Planck scale of the same order of magnitude as that obtained by 
the other methods exposed here.

Consider now 
{\em a matter wave} associated with a particle of mass 
$m$ and Compton wavelength $\lambda 
= h/ (mc)$ scattering off the curvature of spacetime. 
The Planck scale can be pictured as that at which the 
spacetime curvature is caused by the mass $m$ itself and 
the radius of curvature of spacetime due to this mass is 
comparable with the Compton wavelength. Essentially, high 
frequency waves do not backscatter but, at the Planck 
scale, there can be no waves shorter than the background 
curvature radius. Dimensionally, 
the length scale $L$ associated with the mass $m$ (the 
radius of curvature of spacetime) is given by $m =L c^2/G $ 
and quantum and gravitational effects become comparable 
when $\lambda \sim L$, which gives
\be 
\frac{h}{\left( Lc^2/G \right) c}\sim L 
\ee 
or 
\be 
L=\sqrt{\frac{Gh}{c^3}} = \sqrt{2\pi} \, l_\text{pl} \simeq 
2.51 \, l_\text{pl}\,. 
\ee 
In other words, if we pack enough energy into a matter wave 
so that it curves spacetime, the curvature induced by this 
wave 
will impede its own propagation when the Planck scale is 
reached. When the energy of this wave becomes too compact, 
the propagation of the matter wave is affected drastically.

\section{Hawking evaporation of a black hole in a single 
burst}
\label{sec:4}

Hawking's discovery that, quantum mechanically, black holes 
emit a thermal spectrum of radiation allowed for the 
development of black hole thermodynamics by 
assigning a non-zero temperature to black holes.\cite{14} 
In the approximation of a fixed black hole background and of  
a test quantum field in this spacetime, a  
spherical static black hole of mass $m$ emits a thermal 
spectrum at the Hawking temperature
\be\label{HawkingT}
T_\text{H}= \frac{ \hbar c^3}{8\pi G k_\text{B} m} \,.
\ee
As is well known, the emitted radiation peaks at a 
wavelength 
$\lambda_\text{max}$ larger than the 
horizon radius $R_\text{S}=2Gm/c^2$. In fact, using Wien's 
law of 
displacement 
for blackbodies 
\be
\lambda_\text{max}T_\text{H}= b= \frac{h c}{4.9651 
k_\text{B}}\simeq 
2.8978 
\cdot 10^{-3} \, \mbox{m}\cdot \mbox{K}
\ee
and Eq.~(\ref{HawkingT}), one obtains
\be
\lambda_\text{max}= \frac{b}{T_\text{H}}= 
\frac{8\pi^2}{4.9651} \, 
\frac{2Gm}{c^2} \simeq 15.90 R_\text{S} \,.
\ee
Therefore, most of the thermal radiation is emitted at  
wavelengths comparable to, or larger than, the 
black hole horizon, giving a fuzzy image of the black hole.

Heuristically, one can extrapolate Hawking's prediction to a 
Planck regime in which the loss of energy is comparable with the 
black hole mass. Then the Planck scale is reached when the entire 
black hole mass $m$ is radiated in a single burst of  
$N$ particles of wavelength $\sim \lambda_\text{max}$ and 
energy
\be
E =\frac{hc}{\lambda_\text{max}} \sim \frac{hc}{16 
R_\text{S}} 
=\frac{hc^3}{32Gm} \,.
\ee
Although certainly not rigorous, this procedure provides a Planck 
scale of the same order of magnitude as the other procedures 
considered (which is all that one can expect from an order of 
magnitude  estimate).  Assuming $N$ of order unity (say, $N=2$) and  
equating this 
energy with the black hole energy $mc^2$ yields
\be
m \simeq \frac{ NE}{c^2}\simeq  \sqrt{\frac{hc}{16G}} 
=\sqrt{ \frac{\pi}{8}} \, 
m_\text{pl} \sim 0.627 \, m_\text{pl} \,.
\ee

\section{Discussion}
\label{sec:conclusions}

Although black holes are a most striking prediction of 
Einstein's theory of gravity,\cite{12,13} they do not constitute 
the 
entire phenomenology of general relativity and there is no 
need to limit oneself to the black hole concept in 
heuristic derivations of the Planck scale.  
One can consider 
cosmology as well, which is appropriate since cosmology can 
only be discussed in the context of relativistic gravity. 
This approach leads to Barrow's  new heuristic 
derivation of the 
Planck scale \cite{16} by considering, in a 
Friedmann-Lema\^itre-Robertson-Walker universe, a Hubble 
sphere with size comparable to the Compton wavelength of 
the mass it contains. Alternatively, one can consider the
 pair creation of causal bubbles so small that they can be  
treated as 
particles, or one can  derive the Planck scale using the 
scattering of waves off 
the background curvature of spacetime which leads again, in 
order of magnitude, to the Planck scale when applied to 
matter waves. Alternatively, one can consider a black 
hole that evaporates completely in a single burst at the 
Planck 
scale.  Of course, other approaches to the Planck scale are 
in principle conceivable. Although quantum gravity 
is certainly not a subject of undergraduate university 
courses, the 
exercice of imagining new heuristic avenues to the Planck 
scale can be fun and can stimulate the imagination of both 
undergraduate and graduate students, as well as being an 
exercise in dimensional analysis.

\begin{acknowledgments}

The author is grateful to John Barrow for a discussion and 
for pointing out Ref.~8, and to two referees for helpful 
suggestions. This work is supported by Bishop's University 
and by the Natural Sciences and Engineering Research 
Council of Canada.

\end{acknowledgments}

\end{document}